\documentclass{jpp}

\usepackage[squaren]{SIunits}
\usepackage{amsmath}
\usepackage{amssymb}
\usepackage{slashed}
\usepackage{xcolor}
\usepackage{hyperref}

\usepackage{graphicx}

\usepackage{natbib}
\usepackage{lineno}

\DeclareSymbolFont{rsfs}{U}{rsfs}{m}{n}
\DeclareSymbolFontAlphabet{\mathrsfs}{rsfs}

\let\vec\boldsymbol
\renewcommand{\d}{\mathrm d}

\title{Analytical results for non-linear Compton scattering in short intense laser pulses}

\author{Daniel Seipt\aff{1,2}
  \corresp{\email{d.seipt@gsi.de}},
 Vasily Kharin\aff{1},
  Sergey Rykovanov\aff{1},
 Andrey Surzhykov\aff{1},
 \and Stephan Fritzsche\aff{1,2}}

\affiliation{\aff{1}Helmholtz-Institut Jena, Fr{\"o}belstieg 3, 07743 Jena, Germany
\aff{2}
Theoretisch-Physikalisches Institut, Friedrich-Schiller-Universit\"at Jena, Max-Wien-Platz 1, 07743 Jena, Germany}

\begin{document}

\maketitle

\begin{abstract}
We study in detail the strong-field QED process
of non-linear Compton scattering in short intense {plane wave} laser pulses
{of circular polarization}.
Our main focus is placed on how the spectrum of the back-scattered laser light depends
on the shape and duration of the initial short {intense}
pulse.
Although this pulse shape dependence is very complicated and highly non-linear, and has never been addressed explicitly, our analysis reveals that all the dependence on the laser pulse shape is contained in 
a {class of} three-parameter master {integrals}.
Here we present completely analytical expressions for the non-linear Compton spectrum
in terms of 
{these master integrals.}
Moreover, we analyse the universal behaviour of the shape of the spectrum for very high harmonic lines.
\end{abstract}

\section{Introduction}

The non-linear Compton scattering of high-intensity laser pulses off
high-energetic electrons is one of the fundamental processes in strong-field QED.
Its theoretical description goes back to the 1960's where many strong-field QED processes
had been studied in a series of seminal papers
\citep{Nikishov:JETP1964a,Nikishov:JETP1964b,Nikishov:JETP1965,Goldman:PhysLett1964,Brown:PR1964}.
For instance, these authors
predicted the emission of high harmonics and a non-linear intensity-dependent red-shift
of the emitted radiation, which is proportional to $a_0^2$, where $a_0$ is the
dimensionless normalized laser amplitude that is related to the laser intensity ($I$) and wavelength ($\lambda$)  via $a_0^2 = 0.73 \times I[\unit{10^{18}}{\watt\per\centi\metre^2}] \lambda^2[\micro\metre]$.
In a classical picture the generation of high harmonics and the non-linear red-shift
of the emitted radiation can be understood as the influence of the laser's magnetic field due to the $ \vec v\times \vec B$ term in the classical Lorentz force equation.

However, while most contemporary high-intensity laser facilities
generate laser pulses with femtosecond duration
\citep{Mourou:RevModPhys2006,Korzhimanov:PhysUspekh2011,DiPiazza:RevModPhys2012},
most of the early papers on non-linear Compton scattering
did not consider the effect of the finite duration of the
{intense} laser pulse.
For a realistic laser pulse with a finite duration the laser intensity gradually increases from zero to its maximum value. Consequently, the non-linear red-shift is not constant during the
course of the laser pulse and the harmonic lines of the emitted radiation are considerably broadened,
with a large number of spectral lines for each harmonic \citep{Narozhnyi:JETP1996,Boca:PRA2009,
Seipt:PRA2011,Hartemann:PRL1996,Hartemann:PRE1996,Hartemann:PRL2010,Mackenroth:PRA2011,
Dinu:PRA2013}.
In the classical picture this broadening is caused by a gradual slow-down of the
longitudinal electron motion as the laser intensity ramps up \citep{Seipt:PRA2015}.
The occurrence of the additional line structure can be interpreted as interference 
of the radiation that is emitted during different times \citep{Seipt:LasPhys2013}.
The broadening of the spectral lines is especially important with regard to the application of 
non-linear Compton scattering as an x- and gamma-ray radiation source
\citep{Jochmann:PRL2013,Sarri:PRL2014,Rykovanov:JPB2014,Seipt:PRA2015,Khrennikov:PRL2015}.

For ultra-high laser intensities $a_0 \gg 1$ the formation time of the emitted photon is
much shorter than the laser period and the interference of radiation that is emitted
at different times during the course of the pulse is suppressed \citep{Dinu:2015}.
In this regime, where the Compton emission becomes vital for the formation of QED cascades,
the spectrum can be effectively simulated using the photon emission probabilities in a constant crossed field \citep{Ritus:JSLR1985,Fedotov:PRL2010,King:PRA2013,Harvey:PRA2015,Narozhny:Uspekh2015}.
We therefore focus in this paper on the intermediate intensity region $a_0 \sim 1$, where
the interference matters and a general relation between the shape and duration of the laser pulse and the
shape of the spectrum of the backscattered light is very complicated
and highly non-linear.
The shape of the harmonic lines is determined by an interplay between the laser pulse
duration, (ii) spectral composition of the pulse and (iii) the non-linear ponderomotive broadening
which depends on the laser intensity ramps.

In this paper we \textit{analytically} analyse
the non-linear Compton scattering process in a short
{intense plane wave} laser pulse
{of circular polarization}.
{Our analysis is based on the framework of strong-field QED, where the electrons are described as
Volkov states, and we employ the slowly varying envelope approximation.
For convenience, the analysis is performed in incident electron frame of reference.}
In particular, we investigate how the
duration and the shape of the short
{intense} laser pulse
affect the spectrum of the emitted radiation.
We derive a scale invariant master integral that contains
all dependence on the shape of the laser pulse, and we give
explicit analytical expressions for several specific
laser pulse shapes.
Our paper is organized as follows: In Section \ref{sect.theory} we briefly outline the
calculation of the transition amplitude and the energy and angular differential emission probability for non-linear Compton scattering using Volkov states in a pulsed laser field.
The transition amplitude is analysed further in Section \ref{sect.analytic} where we extract the
dependence on the duration and shape of the laser pulse in the form of a master integral.
Explicit analytic expressions for the {master integrals} for different pulse shapes are given in Section
\ref{sect.explicit}.
Throughout the paper we use units with $\hbar=c=1$. Scalar products between four-vectors are
denoted by
$a\cdot b \equiv  a^\mu b_\mu = a^0b^0 - \vec a\vec b$, and the Feynman slash notation is used for
scalar products between four-vectors and the Dirac matrices: $\slashed a \equiv \gamma \cdot a$.

\section{Theoretical Background}

\label{sect.theory}

\begin{figure}
\begin{center}
\includegraphics{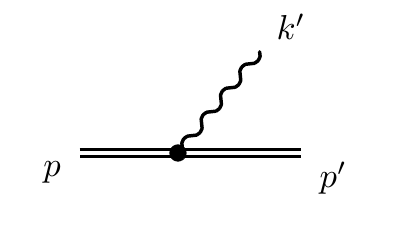}
\end{center}
\caption{Feynman diagram for the emission of a photon with four-momentum $k'$ (wiggly line)
by a laser-dressed Volkov electron with asymptotic four-momentum $p$ (double-line).
After the photon emission the electron has the asymptotic four-momentum $p'$. The double-lines indicate the non-perturbative interaction of the electron with the intense short laser pulse.}
\label{fig:diagram}
\end{figure}

The non-linear Compton scattering process,
i.e.~the emission of a photon by an electron under the action of an intense laser field,
is conveniently described theoretically in  the Furry picture.
The interaction of the electrons with the laser pulse is treated non-perturbatively
by using Volkov electron states $\Psi$ as solutions of the Dirac equation
$(i\slashed \partial -e \slashed  A - m)\Psi = 0$
in the plane-wave background laser field $A$.
Here $m$ and $e=-|e|$ denote the mass and charge of the electron, respectively.
By employing these Volkov states and the strong-field $S$
matrix in the Furry picture can be represented by the Feynman diagram in
Fig.~\ref{fig:diagram}. It is given by the expression
\begin{align}
S = -ie \int \! \d^4 x \, \bar \Psi_{p'}(x) \gamma_\mu {\mathcal A}^\mu_{k'}(x) \Psi_p(x) \,,
\end{align}
where $\mathcal A^\mu_{k'}(x) = (\varepsilon'^*)^\mu e^{ik'\cdot x}$
is the amplitude for the emission of a (non-laser) photon with four momentum $k'$ and polarization $\varepsilon'$, while
$p$ and $p'$ are the asymptotic four-momenta of the electron before and after the photon emission.

In the following, we shall restrict our discussion to the case of a circularly polarized laser pulse with the
four-vector potential in the axial gauge ($k\cdot A=0$)
\begin{align}
A^\mu (\phi) = A_0\, g(\phi)\:  {\Real }\:  \epsilon_+^\mu e^{-i \phi } \,.
\end{align}
It depends only on the phase variable $\phi = k\cdot x$ with the laser photon four-momentum $k=(\omega,0,0,-\omega)$, and with the normalized polarization four-vector
$\epsilon_\pm^\mu = (0,1,\pm i,0)/\sqrt{2}$, with $\epsilon_+ \cdot \epsilon_- = -1$.
The dimensionless normalized laser amplitude is given by $a_0= |e| A_0 /m$.
The shape of the laser pulse is described by an envelope function
$g(\phi)$, that depends on $\phi$ only via the ratio $\phi/\Delta \phi$
with the pulse duration $\Delta \phi$.
Moreover, we use symmetric pulse envelopes with $g(\phi) = g(-\phi)$, $g(0)=1$ and $g(\pm\infty) = 0$.
For the following we shall assume that the laser pulse consists of several optical
cycles such that $\Delta \phi \gg 1$ and we may employ the slowly varying envelope approximation.

One can perform the spatial integrations in the $S$ matrix most conveniently
in light-front coordinates, defined as $x^\pm = x^0 \pm x^3$ and $\vec x^\perp = (x^1,x^2)$, such that
the laser phase is proportional to $\phi = \omega x^+$, and
$\d^4 x =  (2\omega)^{-1} \d \phi \d x^- \d^2 \vec x^\perp $.
After integrating over three light-front coordinates,
the $S$ matrix can be represented in a form \citep{Seipt:PRA2013}
\begin{align}
S = -ie (2\pi)^3 \delta_\mathrm{l.f.}( p - p' - k' ) \, \mathrsfs M(s)
\end{align}
with the light-front delta function
$ \delta_\mathrm{l.f.}( p - p' - k' )
= \frac{1}{\omega}  \delta^2( \vec p_\perp - \vec p_\perp' - \vec k_\perp')  \delta( p^+ - p'^+ -k'^+)$,
enforcing the conservation of three momentum components,
and the transition amplitude
\begin{align}
\mathrsfs  M = \mathrsfs T_0 \mathrsfs C_0 
					+\mathrsfs T_+ \mathrsfs C_+ 
					+\mathrsfs T_- \mathrsfs C_- 
					+\mathrsfs T_2 \mathrsfs C_2 \,.
\label{eq:def.M}
\end{align}
Here, the quantities $\mathrsfs T_j$ denote the transition operators
\begin{align}
\mathrsfs T_0		&=	\bar u_{p'} \slashed \varepsilon'^* u_p \,, \nonumber \\
\mathrsfs T_\pm	&=	\frac{m a_0}{4} \, \bar u_{p'} \left( 
				\frac{\slashed \varepsilon_\pm \slashed k \slashed \varepsilon'^*}{ (k \cdot p')}
				+ \frac{\slashed \varepsilon'^* \slashed k \slashed \varepsilon_\pm }{(k\cdot p)}
			\right) u_p \,, \\
\mathrsfs T_2		&= \frac{m^2 a_0^2 \, (\varepsilon'^* \cdot k )}{4 (k\cdot p) (k\cdot p')} \, \bar u_{p'} \slashed k u_p \,, \nonumber
\end{align}
which are sensitive to the spin of the incident and final electrons (via the spinors $u_p$ and $\bar u_{p'}$)
and the polarization of the emitted photon. However, they are only weakly dependent on the energy and momentum of the emitted photon.
The dependence on the dynamics of the scattering process is mainly contained in the so-called dynamic
integrals over the laser phase
\begin{align}
\mathrsfs C_{\pm 1} & 
		= \intop_{-\infty}^\infty \! \d \phi \,g(\phi) e^{\mp i \phi} 
		\exp \left\{ 
		i \ell \left[  \phi + \alpha g(\phi) \sin(\phi+\varphi) + \beta \int_0^\phi \! \d \phi' \, g^2(\phi')  \right]
		\right\} \,, \nonumber
	\\
\mathrsfs C_2    & 
		= \intop_{-\infty}^\infty  \! \d \phi \,g^2(\phi) 
		\exp \left\{ 
		i \ell \left[  \phi + \alpha g(\phi) \sin(\phi+\varphi) + \beta \int_0^\phi \! \d \phi' \, g^2(\phi')  \right]
		\right\}  \,.
			\label{eq:def.C2}
\end{align}
Here we employed the slowly varying envelope approximation for the laser pulse
and we use the definition $\varphi = \arctan k'_y/k'_x$.
The fourth dynamic integral $\mathrsfs C_0$ is represented as a combination of the other three integrals, defined in Eqs.~\eqref{eq:def.C2}, as
\begin{align}
\mathrsfs C_0 = - \frac{\alpha}{2} 
			\left( e^{-i\varphi} \mathrsfs C_+  + e^{i\varphi} \mathrsfs C_- \right)
			 - \beta \mathrsfs C_2 \,,
\end{align}
by the requirement of the gauge invariance of the $S$ matrix \citep{Ilderton:PRL2011,diss:Seipt}.

Here we have defined $\ell$ as the amount of four-momentum that is absorbed from the laser field
\begin{align}
\ell  \equiv \frac{k'\cdot p}{k\cdot p'} = \frac{p'^- + k'^- - p^-}{k^-}\,,
\end{align}
and provides a Lorentz-invariant way to parametrise the frequency of the emitted photon
\begin{align}
\omega'( \ell ) = \frac{\ell \omega}{1 + \ell \frac{\omega}{m} (1+\cos \vartheta)} \,.
\label{eq:def.omega}
\end{align}
For convenience, we moved to the rest-frame of the incident electron,
where $p = (m,0,0,0)$.
Moreover, we defined the coefficients
\begin{align}
\alpha &=   \frac{a_0}{\sqrt2} \sin \vartheta \,,
	\label{eq:def.alpha} \\
\beta &=  \frac{a_0^2}{4} (1 + \cos \vartheta) \,.
	\label{eq:def.beta}
\end{align}
Using the the transition amplitude \eqref{eq:def.M},
the angular- and energy-differential photon emission probability
is given by \citep{Seipt:PRA2011,diss:Seipt}
\begin{align}
\frac{\d W}{\d \omega' \d \Omega} = \frac{e^2 \omega' |\mathrsfs M|^2}{64 \pi^3 (k\cdot p) (k\cdot p')} \,.
\label{eq:spectrum}
\end{align}
In order to study how the differential emission probability depends on the laser pulse parameters---%
the laser strength $a_0$, pulse duration $\Delta \phi$, or the shape of the pulse envelope $g$---%
it is required to evaluate the dynamic integrals $\mathrsfs C_j$.

For many purposes it is sufficient to perform the integrations over the laser phase numerically,
as was done for instance in Refs.~\citep{Seipt:PRA2011,Mackenroth:PRA2011,Krajewska:PRA2012,Twardy:JPCS2014}.
Another approach to calculate the spectrum \eqref{eq:spectrum}
relies on a saddle point analysis of the highly oscillating
phase integrals in \eqref{eq:def.C2}
\citep{Narozhnyi:JETP1996,Seipt:LasPhys2013,Mackenroth:PRL2010,Seipt:PRA2015,Seipt:2015}.
A third possibility to evaluate the dynamic integrals would be an attempt to find analytic solutions.
Such a completely analytical evaluation has been done for instance
in \citep{Hartemann:PRE1996} for the on-axis radiation spectrum
{using classical electrodynamics}.
This approach will be pursued in this paper for arbitrary emission angles and general pulse shapes.
In the following we will present a completely analytical evaluation of the dynamic integrals
to gain more insight on the dependence of the spectrum of the backscattered light on the
laser pulse duration and envelope shape.

\section{Analytic Evaluation of the Dynamic Integrals}
\label{sect.analytic}

In this section we further analyse the properties of the spectrum of the emitted radiation.
We apply a detailed mathematical analysis to the transition amplitude $\mathrsfs M$,
and in particular the dynamic integrals $\mathrsfs C_j$,
from the previous section in order to extract how
they depend on the laser pulse duration and pulse shape.
Here, we aim to provide explicit analytic expressions for the dynamic integrals
\eqref{eq:def.C2}.

It is known from previous studies
\citep{Narozhnyi:JETP1996,Seipt:LasPhys2013,Seipt:PRA2015}
that the oscillating term ($\propto \alpha$) in the exponent of the 
dynamic integrals, Eq.~\eqref{eq:def.C2}, is responsible for the emission of high harmonics.
The term containing the integral over the squared pulse envelope
($\propto \beta$) changes only slowly as a function of $\phi$.
This so-called ponderomotive term is responsible for the broadening of the harmonic lines and
the spectral structures seen within each harmonic.
In the following, we first disentangle
these two effects (Section \ref{sect.analytic.exp})
and later on analyse the ponderomotive broadening for each harmonic line (Sections \ref{sect.analytic.reduction} et seq.)

\subsection{Expansion into Harmonics}
\label{sect.analytic.exp}

Let us first expand the dynamic integrals into a sum of partial terms which
can be interpreted as the emission of higher harmonics in analogy to the case of infinite plane waves.
Following~\cite{Narozhnyi:JETP1996} and \cite{Seipt:LasPhys2013},
we define a generalized floating-window Fourier series for a non-periodic function $f(\phi)$
\begin{align}
f(\phi) & = \sum_{n = -\infty}^\infty c_n(\phi) e^{ - i n \phi } \,, \qquad \qquad \qquad
 c_n(\phi) = \frac{1}{2\pi} \, \intop_{\phi-\pi}^{\phi+\pi} \!\d\phi' \, f(\phi') \, e^{i n \phi' } 
\end{align}
with Fourier coefficients
$ c_n(\phi) $
that depend on the location of the window centre.
Applying this floating window Fourier series to the integrand of the dynamic integrals,
and using the slowly varying envelope approximation \citep{Narozhnyi:JETP1996},
yields a generalized Jacobi-Anger type expansion
\begin{align}
e^{i \ell \alpha g(\phi) \sin{ ( \phi+\varphi )}} = \sum_{n} (-1)^n 
			\, J_n(\ell \alpha g(\phi) ) \: e^{-in(\phi+\varphi)} \,,
			\label{eq:Jacobi:Anger}
\end{align}
with the Bessel function of the first kind $J_n(z)$ \citep{book:Watson}.
This expansion strongly resembles the expansion into harmonics known from the
well-studied case of infinite plane waves \citep{book:Landau4}, where $g = 1$.
Note, however, that here the argument of the Bessel functions depends on the laser phase $\phi$ via
laser pulse envelope $g(\phi)$.

Employing the above expansion we can cast the dynamic integrals into a form
\begin{align}
\mathrsfs C_2(\ell ) &= \sum_n (-1)^n \,  e^{-in\varphi} \: 
			C_2^{(n)} (\ell ) \,, \nonumber \\
\mathrsfs C_\pm(\ell ) &= \sum_n (-1)^n \, e^{-i(n\mp1)\varphi} \: 
			C_\pm^{(n)} (\ell) \label{eq:C1.harmonics}\,,
\end{align}
with
\begin{align}
C_2^{(n)}(\ell ) 		&= \intop_{-\infty}^\infty \! \d \phi \:
						g^2(\phi) J_n(\ell \alpha g) e^{i(\ell-n)\phi + i\ell\beta \int \!\d \phi \, g^2} \,, \nonumber \\
C_\pm^{(n)}(\ell ) &=  \intop_{-\infty}^\infty \! \d \phi \:
						g(\phi) J_{n\mp 1}(\ell  \alpha g) e^{i(\ell-n)\phi + i\ell\beta \int \!\d \phi \, g^2} \,.
\end{align}
Note that for symmetric laser pulse envelopes, as we use in this paper,
all the coefficients $C_j^{(n)}(\ell )$ are purely real-valued.
Making use of the expansions~\eqref{eq:C1.harmonics}, the transition amplitude $\mathrsfs M$
can be written as a sum of partial amplitudes {$\mathrsfs M^{(n)}$
via} $\mathrsfs M = \sum_{n=1}^\infty \mathrsfs M^{(n)}$, representing the emission of the $n$-th harmonic.
The shape of each of the harmonic lines is determined by the integrals $C_j^{(n)}(\ell)$,
which might be called the partial dynamic integrals for the $n$-th harmonic.
Unfortunately, in these integrals the pulse envelope $g$ appears as the argument of the Bessel function,
preventing their immediate analytic evaluation.

The parameter $\alpha$ that appears in the argument of the Bessel functions
goes to zero for on-axis radiation, $\vartheta = 0$.
Since the Bessel functions behave as $J_n(z) \approx \frac{z^n}{2^n n!}$ for small argument $z$ this means that only the first harmonic $n=1$ is emitted on-axis for a circularly polarized laser pulse,
with the only contribution coming from $C_+^{(1)}$.
The result that no higher harmonics occur on-axis 
is known from the case of infinitely long plane waves as ``blind spot'' or
``dead cone'' in the literature (see e.g.~\cite{Harvey:PRA2009} and references therein).

\subsection{Reduction to a Master Integral}
\label{sect.analytic.reduction}

The next important step is to extract the pulse shape function $g$ from the argument
of the Bessel functions in the definition of the integrals $C_j^{(n)}$. This will eventually
allow to define a master integral that contains all the dependence on the laser pulse envelope.
Such an extraction is achieved by
applying the multiple argument expansion for the Bessel functions \citep{book:Watson}
\begin{align}
J_n\left( \ell \alpha g(\phi) \right) = g^n(\phi) \sum_{k=0}^\infty \left[ 1-g^2(\phi) \right]^k 
	\frac{J_{n+k}(\ell \alpha)}{k!} 
	\left( \frac{\ell \alpha}{2}\right)^k \,.
	\label{eq:BesselJ.multi}
\end{align}
Thus, instead of having to deal with the pulse envelope $g$ as an argument of the $n$-th Bessel function
we now get a power series in $(1-g^2)$, with the coefficients containing higher-order Bessel functions.
We should note that the overlap of the functions $g^n$ and some power of $(1-g^2)^k$ rapidly gets small
for increasing $n$ and $k$.
The powers of $g^n$ are localized at the origin $\phi=0$ more strongly for larger values of $n$, while the powers of $(1-g^2)$ vanish at the origin.
Their product in the expansion \eqref{eq:BesselJ.multi} samples the edges of the laser pulse.

Employing the above expansions we obtain for the partial dynamic integrals for the $n$-th harmonic
the series
\begin{align}
C_2^{(n)}(\ell)		&= \sum_{k=0}^\infty 	\frac{J_{n+k}(\ell \alpha)}{k!}  
						\left( \frac{\ell\alpha}{2}\right)^k
						B^k_{n+2} (\ell-n,\ell\beta) \,, \\
C_\pm^{(n)}(\ell) &= \sum_{k=0}^\infty 	\frac{J_{n+k\mp 1}(\ell \alpha)}{k!} 
						 \left( \frac{\ell\alpha}{2}\right)^k
						B^k_{n+1\mp1} (\ell-n,\ell \beta) \,,
\end{align}
where we have defined the
ponderomotive integrals
\begin{align}
B^k_r(\ell-n , \ell\beta) &=
\intop_{-\infty}^\infty \! \d \phi \, 
						g^{r}(\phi) \left[1-g^2(\phi)\right]^k \, e^{i( \ell -n)\phi + i \ell \beta \int \!\d \phi \, g^2} \,.
						\label{eq:def.Bkr}
\end{align}
They contain all dependence on the laser pulse shape and pulse duration
and its influence on the longitudinal electron motion and spectral broadening.

Before evaluating these ponderomotive integrals further,
let us first discuss the limit of infinite plane waves, $g\to1$,
where the laser intensity is switched on adiabatically at past infinity
and then stays constant.
As a consequence there is no ponderomotive broadening for
the infinite plane waves.
Because of $1-g^2 = 0$ we find
\begin{align}
B^k_r(\ell-n,\ell\beta) &\stackrel{g=1}{\longrightarrow} 
		\delta_{k0}  \int \! \d \phi \, e^{i(\ell - n)\phi + i \ell \beta \phi} 
		= 2\pi \, \delta_{k0} \, \delta(\ell - n +\ell \beta)\,.
\end{align}
The delta function here restricts the generally continuous variable $\ell$ to discrete values
$\ell_n = n/(1+\beta)$.
Thus, the frequency of the emitted photon, Eq.~\eqref{eq:def.omega}, becomes discrete as well:
\begin{align}
\omega'_n \equiv \omega'(\ell_n) = \frac{n \omega}{1 + \beta + \frac{n \omega}{m} (1+\cos \vartheta)} 
	= \frac{n \omega}{1 + \left( \frac{n \omega}{m} + \frac{a_0^2}{4}\right)(1+\cos \vartheta)} 
	\label{eq:omegan}
\end{align}
with the well-known non-linear intensity dependent red-shift \citep{book:Landau4},
but no spectral broadening.
Eq.~\eqref{eq:omegan} is usually interpreted as the absorption of $n$ laser photons and the emission
of high harmonics.
Moreover, in the expansion \eqref{eq:BesselJ.multi} all terms with $k>0$ vanish and we re-obtain
the well known result that the partial matrix element of the $n$-th harmonic contains the
Bessel functions $J_n$ and $J_{n\pm1}$ \citep{book:Landau4,Ritus:JSLR1985}.
 
It is possible to find a recurrence relation for the ponderomotive integrals:
\begin{align}
B^k_r = B^{k-1}_r - B^{k-1}_{r+2}\,.
\end{align}
Subsequent application of this relation helps to reduce the order of the upper index to zero:
\begin{align}
B^k_r = \sum_{\nu=0}^k (-1)^\nu \binom{k}{\nu} B^0_{r+2\nu} \,. 
\end{align}
Thus, we have to calculate only those ponderomotive integrals with the upper index $k=0$.
Let us now re-scale the integration variable
in \eqref{eq:def.Bkr} as $\phi \to t =  \phi/\Delta \phi $, in order to define the three-parameter
{master integrals} as
\begin{align}
 \mathcal B_r( \xi , \eta) \: \equiv \:  \frac{B_r^0(\ell - n ,\ell \beta)}{\Delta \phi} \,,
\end{align}
as a function of the rescaled variables 
{
\begin{align}
\xi &= (\ell-n) \Delta \phi \,,\\
 \eta &= \ell\beta \Delta \phi \,,
\end{align}}
and for positive integer values of $r$.
{Note that the variable $\xi$ depends on the harmonic number $n$.}
{
An explicit relation between the physically accessible variables $(\omega',\vartheta)$
and the abstract variables $(\xi,\eta)$ is given at the end of this subsection in Eqs.~\eqref{eq:unfold1}--\eqref{eq:unfold2}.}
%
The {master integrals explicitly read}
\begin{align}
\mathcal B_r (\xi,\eta) \equiv \intop_{-\infty}^\infty \! \d t \, g^r(t) \, e^{i \xi t + i \eta \int \! \d t \, g^2(t) }
\,.
\label{eq:def.master.integral}
\end{align}
It only depends on the shape of the laser pulse and is completely independent of the pulse duration.
Note that the master integrals are real-valued functions for all symmetric laser pulse shapes.

\begin{figure}
\begin{center}
\includegraphics[width=0.5\textwidth]{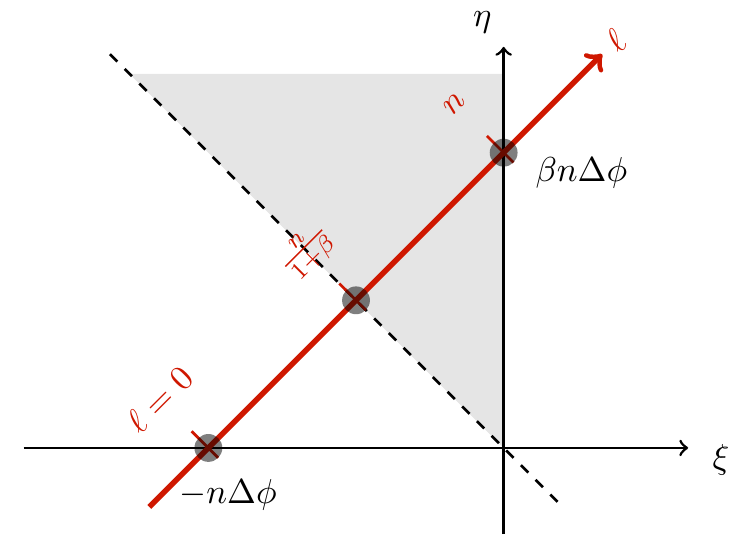}
\end{center}
\caption{Illustration how to cut the $\xi$--$\eta$ plane in order to obtain
the shape of the spectral lines as a function of $\ell$ {for fixed scattering angle $\vartheta$}.}
\label{fig:cutting}
\end{figure}

From stationary phase arguments one can deduce that the master integrals are essentially different
from zero only in the regions bounded by the $\eta$--axis and the line $\eta = -\xi$.
In this region the {master integrals are} oscillating function for all values of $r$.
This region is visualized as a grey shaded area in Fig.~\ref{fig:cutting}.
Outside of this region it rapidly approaches zero.

Before we continue our discussion of the properties of the {master integrals
and their} pulse shape dependence, let us first represent the partial transition amplitudes
$\mathrsfs M^{(n)}$ in terms of the $\mathcal B_r(\xi,\eta) $ by
putting together all the expansions:
\begin{align}
\mathrsfs M^{(n)} &= \Delta \phi (-1)^n e^{-in\varphi}
\sum_{k=0}^\infty \sum_{\nu=0}^k \frac{(-1)^\nu 	}{k!}  \binom{k}{\nu} 
			\left( \frac{\ell \alpha}{2} \right)^k \nonumber \\
			&\qquad \times 		\bigg\{
			e^{i\varphi} J_{n+k-1}(\ell \alpha) 
			\left[ \mathrsfs T_+ - \frac{\alpha e^{-i\phi}}{2}\mathrsfs T_0 \right] \mathcal B_{n+2\nu} (\xi,\eta)
			\nonumber \\
			& \qquad \qquad +
			e^{i\varphi} J_{n+k+1}(\ell \alpha) 
			\left[ \mathrsfs T_- - \frac{\alpha e^{+i\phi}}{2}\mathrsfs T_0 \right] \mathcal B_{n+2+2\nu} (\xi,\eta)\nonumber \\
			& \qquad \qquad +
            J_{n+k}(\ell \alpha) 
			\Big[ \mathrsfs T_2 - \beta \mathrsfs T_0 \Big]\, \mathcal B_{n+2+2\nu} (\xi,\eta)
			\bigg\}
			\,. \label{eq:partial:M:PPW}
\end{align}
To obtain the spectrum of non-linear Compton scattering we have to plug this transition amplitude
into Eq.~\eqref{eq:spectrum}.

For the sake of completeness let us now briefly discuss the partial transition amplitudes in the
limit of infinite plane wave laser fields, $g=1$.
In this case the {master integrals} turns into
$\mathcal B_r( \xi,\eta) \to 2\pi \delta( \xi +\eta )$, i.e.~they
are localized along the diagonal $\eta = - \xi$.
Moreover, the summation over $\nu$ yields just a Kronecker delta $\delta_{k0}$ such that only the $k=0$ term survives in the sum over $k$.
Therefore we obtain for the transition amplitude for infinite plane waves
\begin{align}
\mathrsfs M^{(n)} &= 2\pi \delta( \ell-n+ \ell \beta ) \Delta \phi (-1)^n e^{-in\varphi}
				\: \bigg\{
			e^{i\varphi} J_{n-1}(\ell_n \alpha) 
			\left[ \mathrsfs T_+ - \frac{\alpha e^{-i\phi}}{2}\mathrsfs T_0 \right]
			\nonumber \\
			& \qquad \qquad \quad  +
			e^{i\varphi} J_{n+1}(\ell_n \alpha) 
			\left[ \mathrsfs T_- - \frac{\alpha e^{+i\phi}}{2}\mathrsfs T_0 \right]  +
            J_{n}(\ell_n \alpha) 
			\Big[ \mathrsfs T_2 - \beta \mathrsfs T_0 \Big]
			\bigg\}
			\,, \label{eq:partial:M:IPW}
\end{align}
with the argument of the Bessel functions now being $\ell_n\alpha = n\alpha / (1+\beta)$, which
reproduces the well-known textbook result \citep{book:Landau4}.
The direct comparison between \eqref{eq:partial:M:PPW} and \eqref{eq:partial:M:IPW} impressively
demonstrates how much more complex and intricate the case of the pulsed laser fields is, as compared
to infinite plane waves.
The master integrals $\mathcal B_r(\xi,\eta)$, which become
trivial in the case of infinite plane waves,
cause the increased complexity of the transition amplitude for pulsed plane wave laser fields.
They can be considered as a fingerprint of the laser pulse shape.

Let us now return to our discussion of the properties of the {master integrals} Eq.~\eqref{eq:def.Bkr}
by first discussing how the rescaled arguments $\xi$ and $\eta$ relate to the frequency $\omega'$ and 
scattering angle $\vartheta$ of the emitted photon.
In order to obtain the shape of the frequency spectrum as a function of $\ell$
(or the photon frequency $\omega'$ by means of Eq.~\eqref{eq:def.omega}) one has to
cut the functions $\mathcal B_r(\xi ,\eta)$ along the straight line $\eta = \beta \xi + n \beta \Delta \phi $
in the $\xi$--$\eta$ plane, depicted as a red diagonal line in Fig.~\ref{fig:cutting}.
This line intersects the $\xi$--axis at $\xi=-n\Delta\phi$, the slope is just $\beta$ (i.e.~it depends on the laser intensity), and the
intersection with the $\eta$--axis is at $\eta =  \beta n \Delta \phi$. 
Note that the dependence on the scattering angle $\vartheta$ is entirely contained in the parameter
$\beta$, see Eq.~\eqref{eq:def.beta}.

It is important to note which values of $\ell$ lie inside the grey shaded area where the
{master integrals are}
non-zero: They are exactly those values between the red-shifted and unshifted $n$-th harmonic
lines in the infinite plane wave: $n/(1+\beta) \leq \ell \leq n$, see Fig.~\ref{fig:cutting}.
Moreover, we see that the diagonal $\eta=-\xi$ represents the red-shifted harmonics in the infinite monochromatic plane wave.
That means, the region close to the diagonal line $\eta = -\xi$ is formed close to the centre of the laser pulse where the intensity is largest. In the region close to the $\eta$--axis the {master integrals} is
formed at the very edges of the laser pulse where the intensity is very low.

\begin{figure}
\begin{center}
\includegraphics[width=0.32\columnwidth]{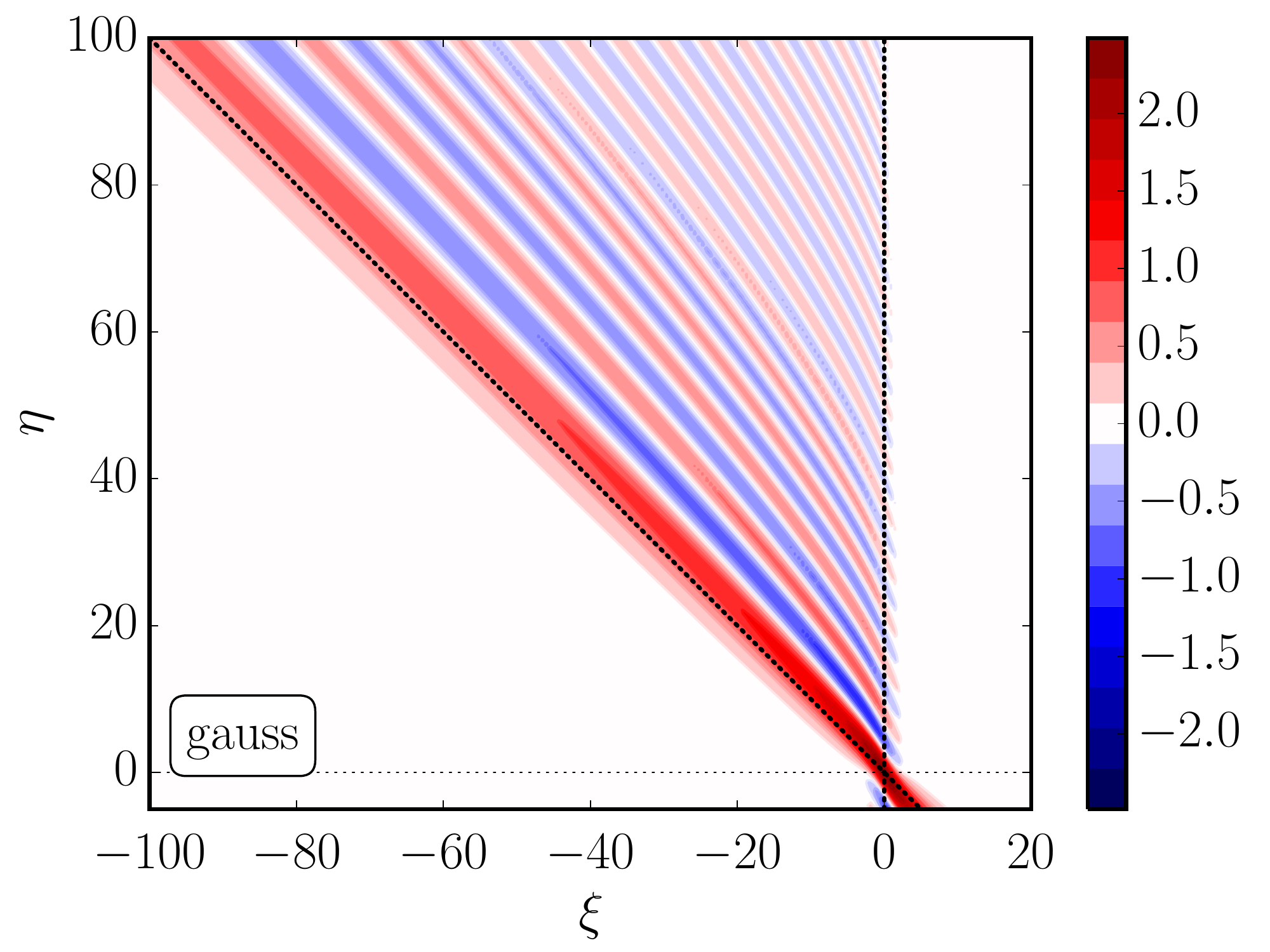}
\includegraphics[width=0.32\columnwidth]{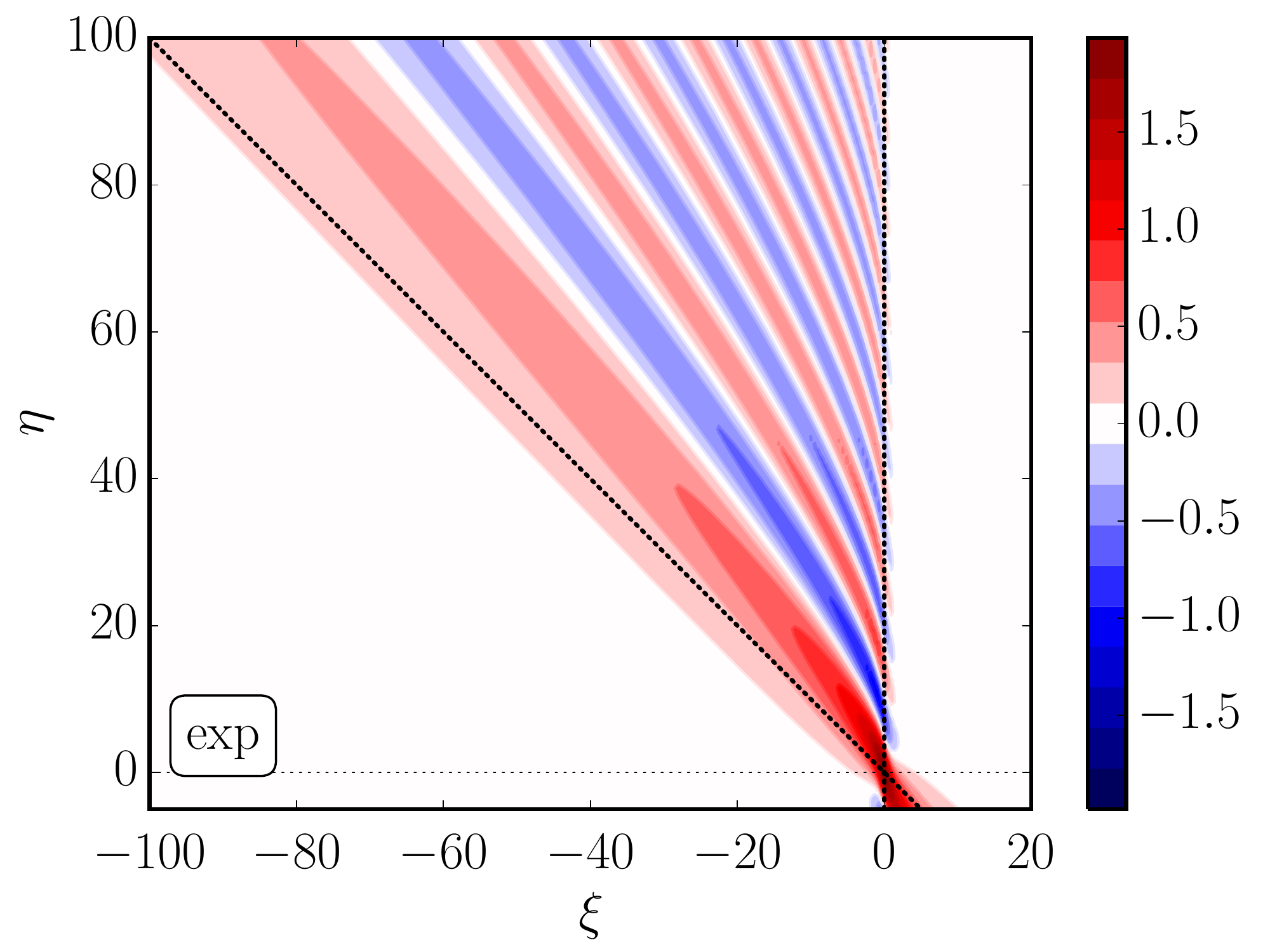}
\includegraphics[width=0.32\columnwidth]{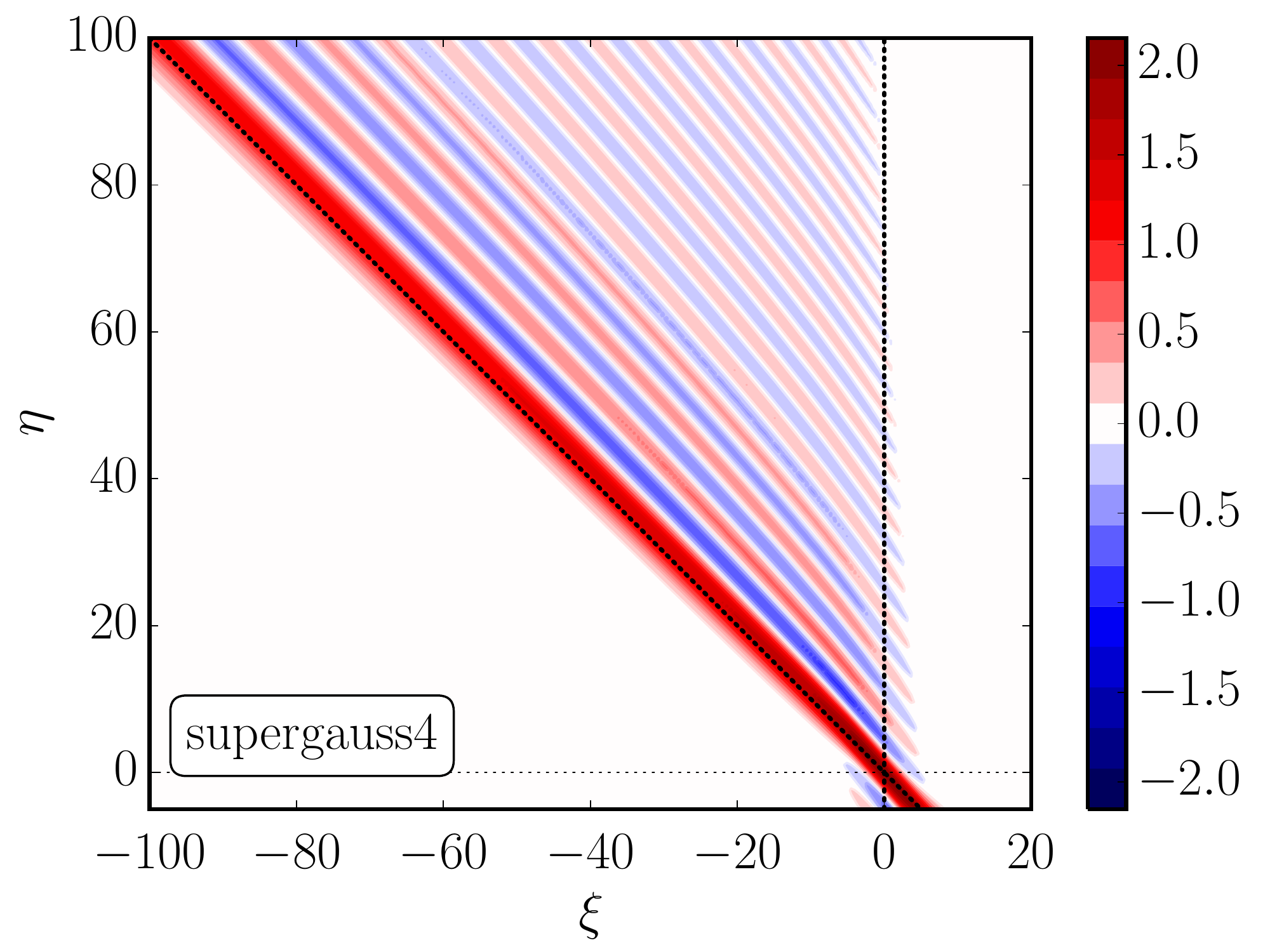}
\end{center}
\caption{{Numerical} evaluation of the master integral $\mathcal B_1(\xi, \eta)$
for for different pulse shapes: a Gaussian $g(t)=e^{-t^2/2}$ (left), an exponential $g=e^{-|t|}$ (centre),
and a Supergaussian $g = e^{-t^4/2}$ (right).
\label{fig:shapes}}
\end{figure}

\begin{figure}
\begin{center}
\includegraphics[width=0.32\columnwidth]{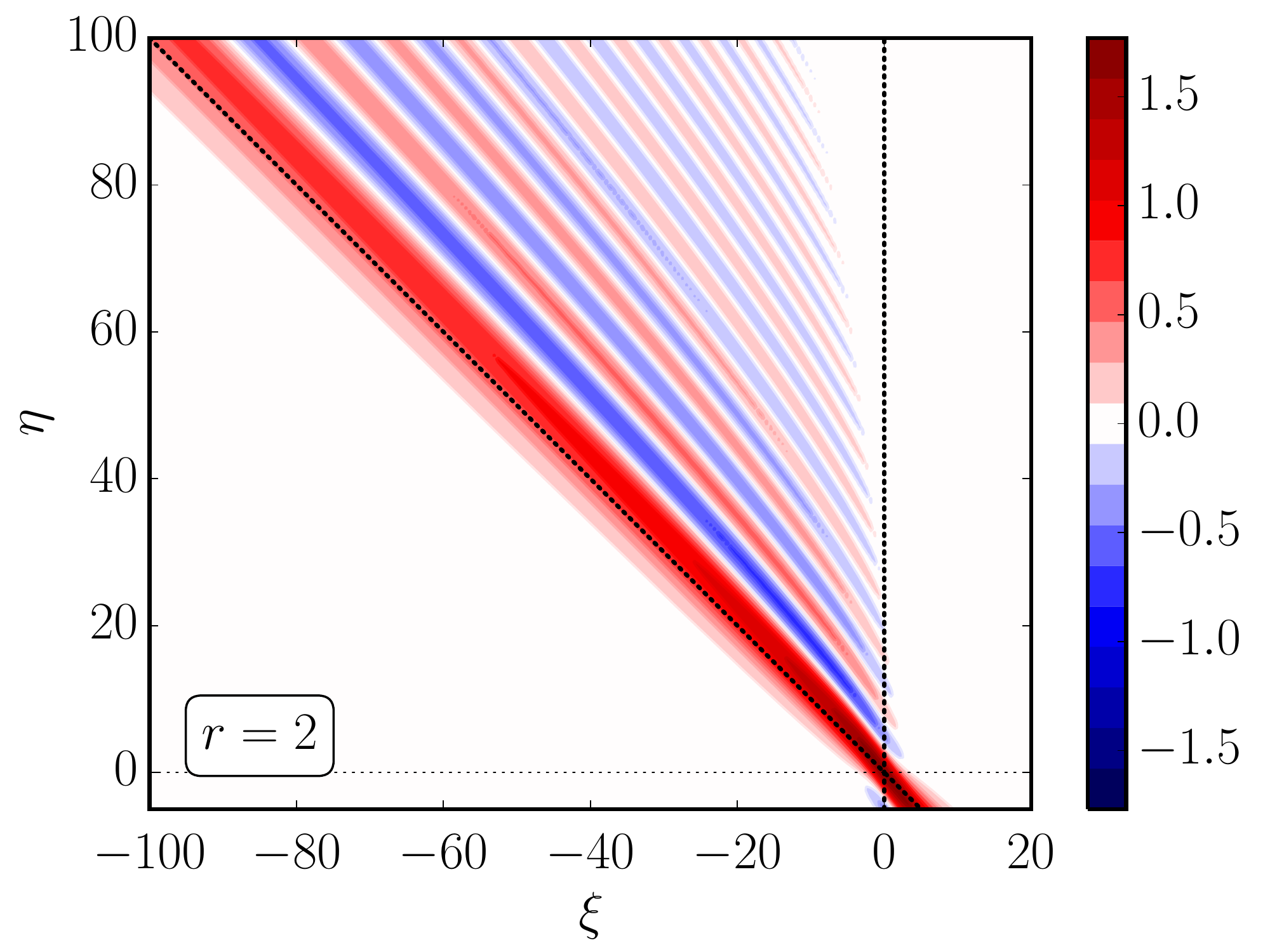}
\includegraphics[width=0.32\columnwidth]{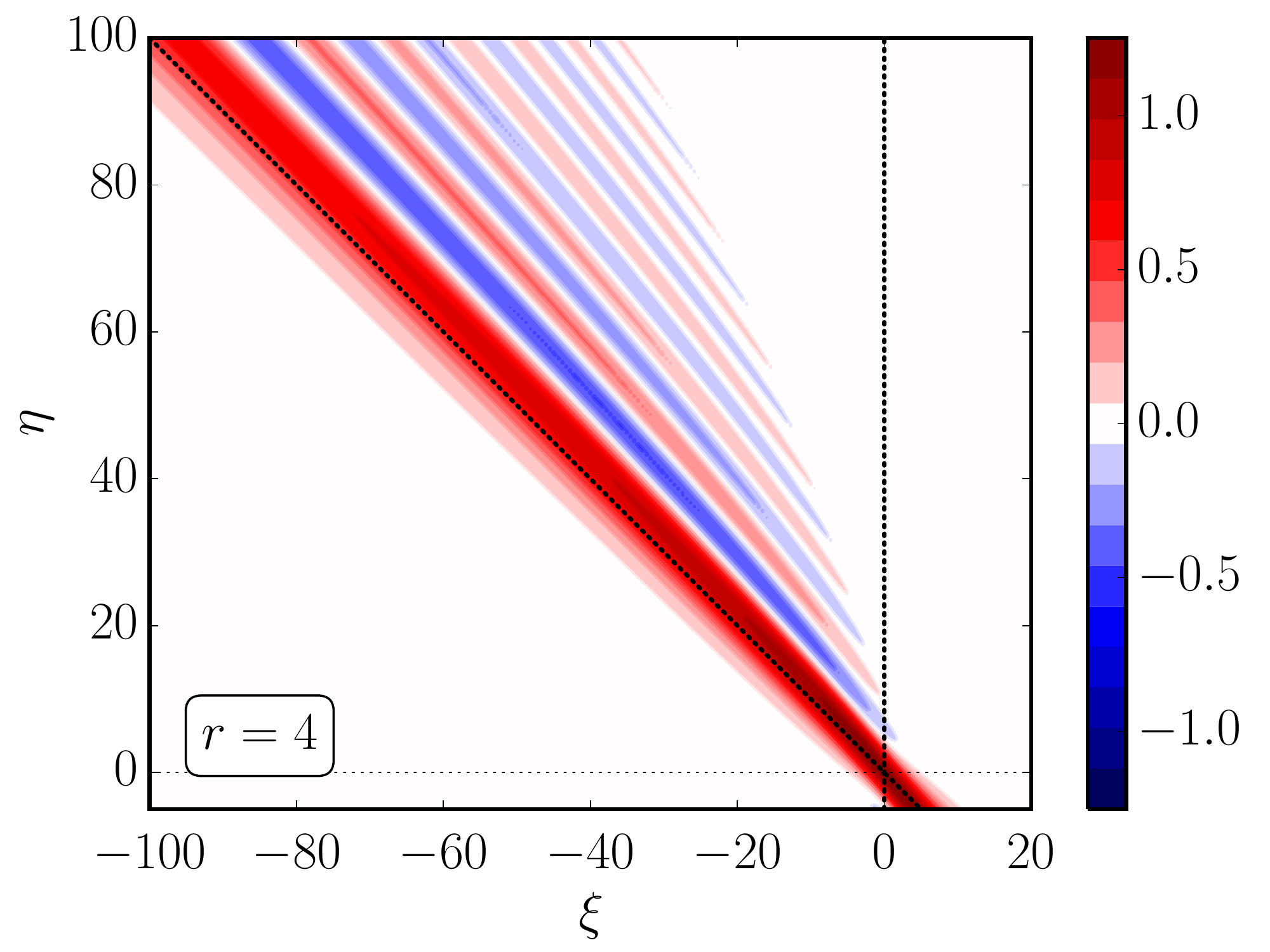}
\includegraphics[width=0.32\columnwidth]{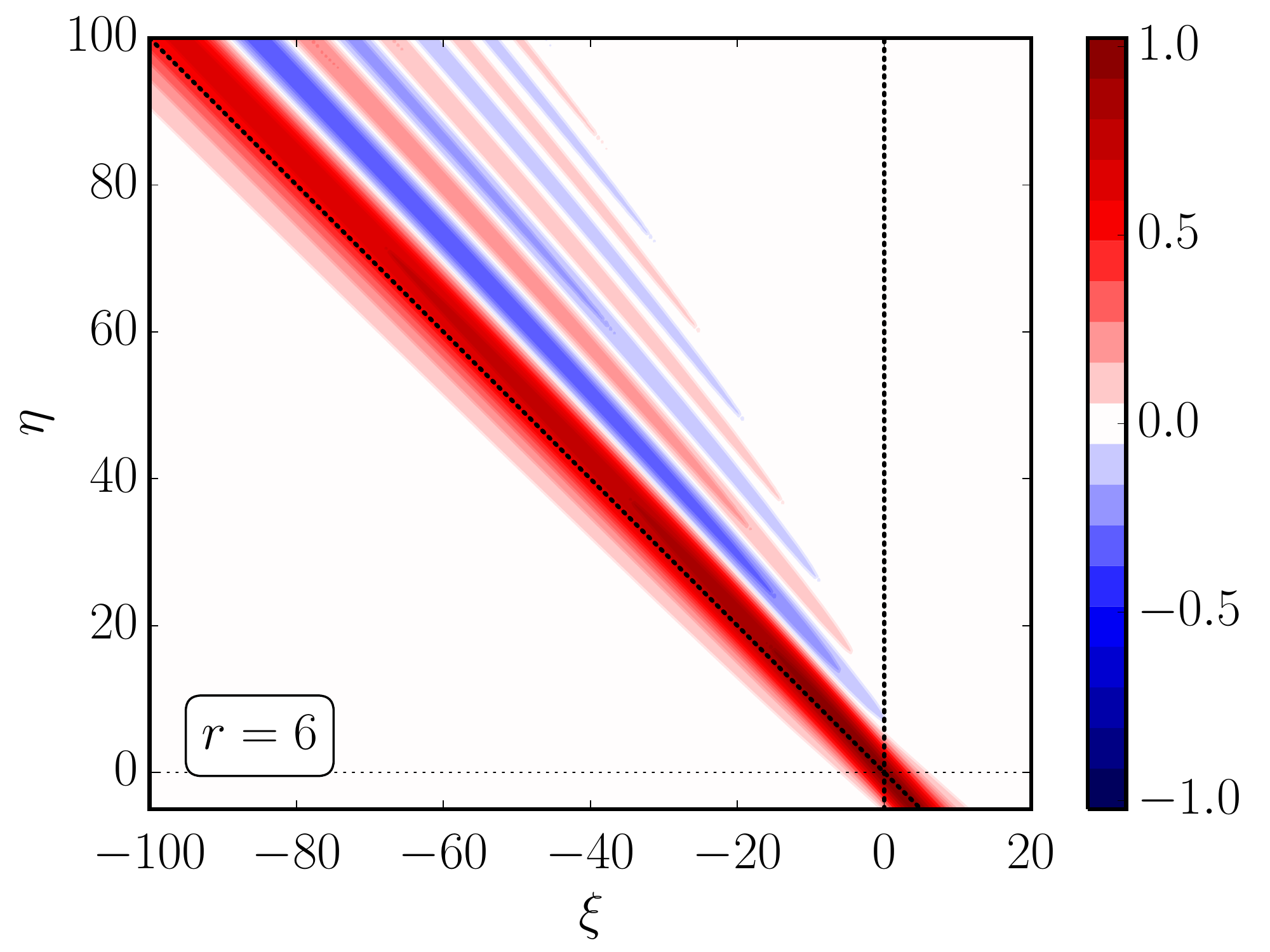}
\end{center}
\caption{{Numerical} evaluation of the master integral \eqref{eq:def.master.integral}
for a Gaussian pulse envelope $g=e^{-t^2/2}$ in the
$\xi$--$\eta$ plane for increasing values of $r$ from left to right.
\label{fig:gauss}}
\end{figure}

{Numerical} evaluations of the master integral $\mathcal B_1(\xi, \eta)$ are depicted in Fig.~\ref{fig:shapes}
for three different pulse envelopes. We see that each pulse shape generates a distinct pattern
of oscillations in the triangular region bounded by the $\eta$--axis and the diagonal $\eta=-\xi$.
For a Gaussian pulse envelope $g(t) = e^{-t^2/2}$ {numerical}
evaluations of the master integrals are depicted in Fig.~\ref{fig:gauss}
for different values of $r$.
One can see that the larger the value of $r$ the stronger the function is localized close to the line $\eta = -\xi$ (i.e.~the non-linear Compton edge in
the limit of infinite plane waves).
By recalling how we need to cut the $\xi$--$\eta$ plane to obtain the frequency spectrum
we easily deduce that for longer pulses or higher harmonics the spectral lines contain more oscillations.
This observation is in line with results using the saddle point method \citep{Seipt:LasPhys2013}.

\begin{figure}
\begin{center}
\includegraphics[width=0.32\columnwidth]{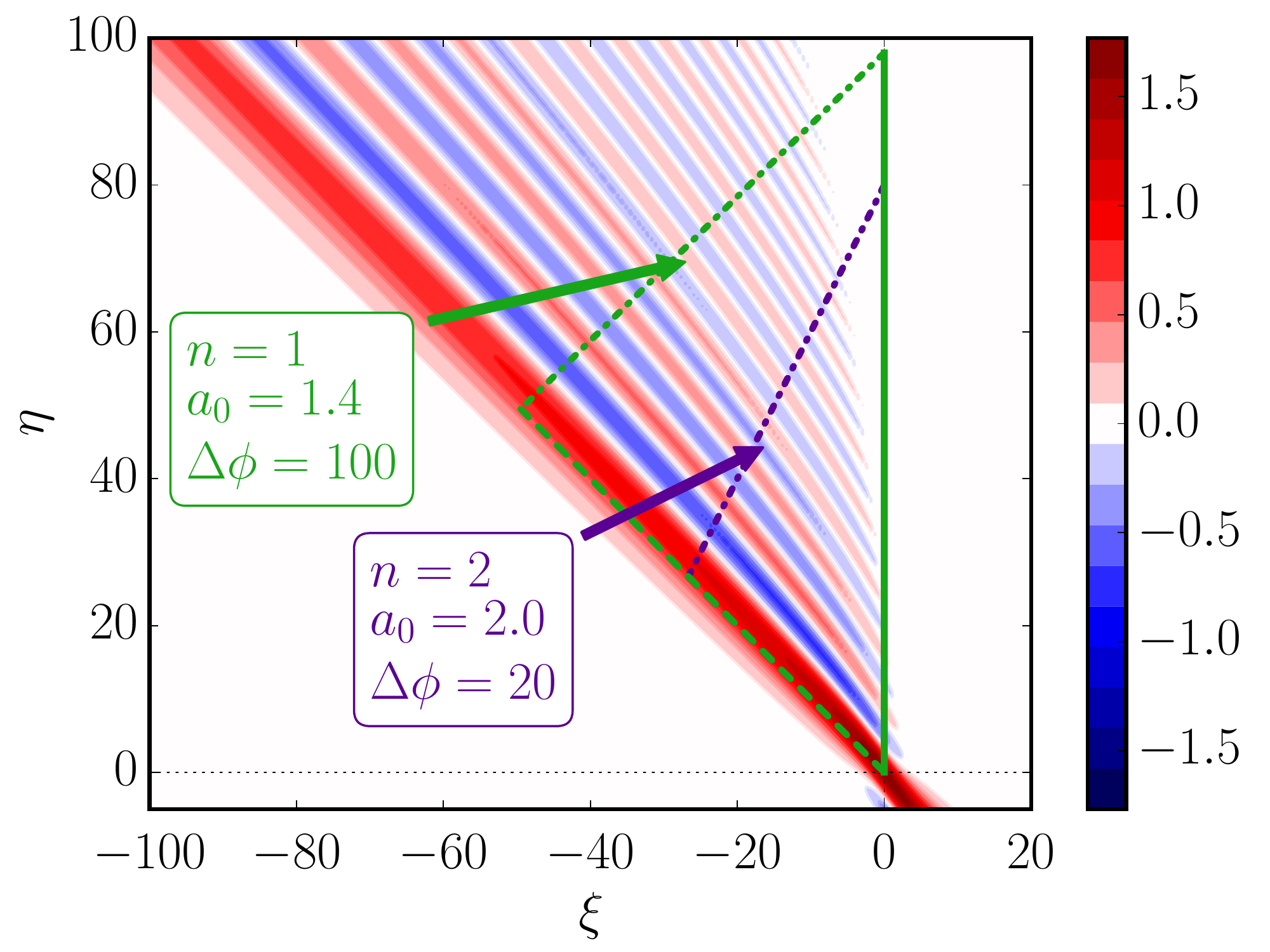}
\includegraphics[width=0.32\columnwidth]{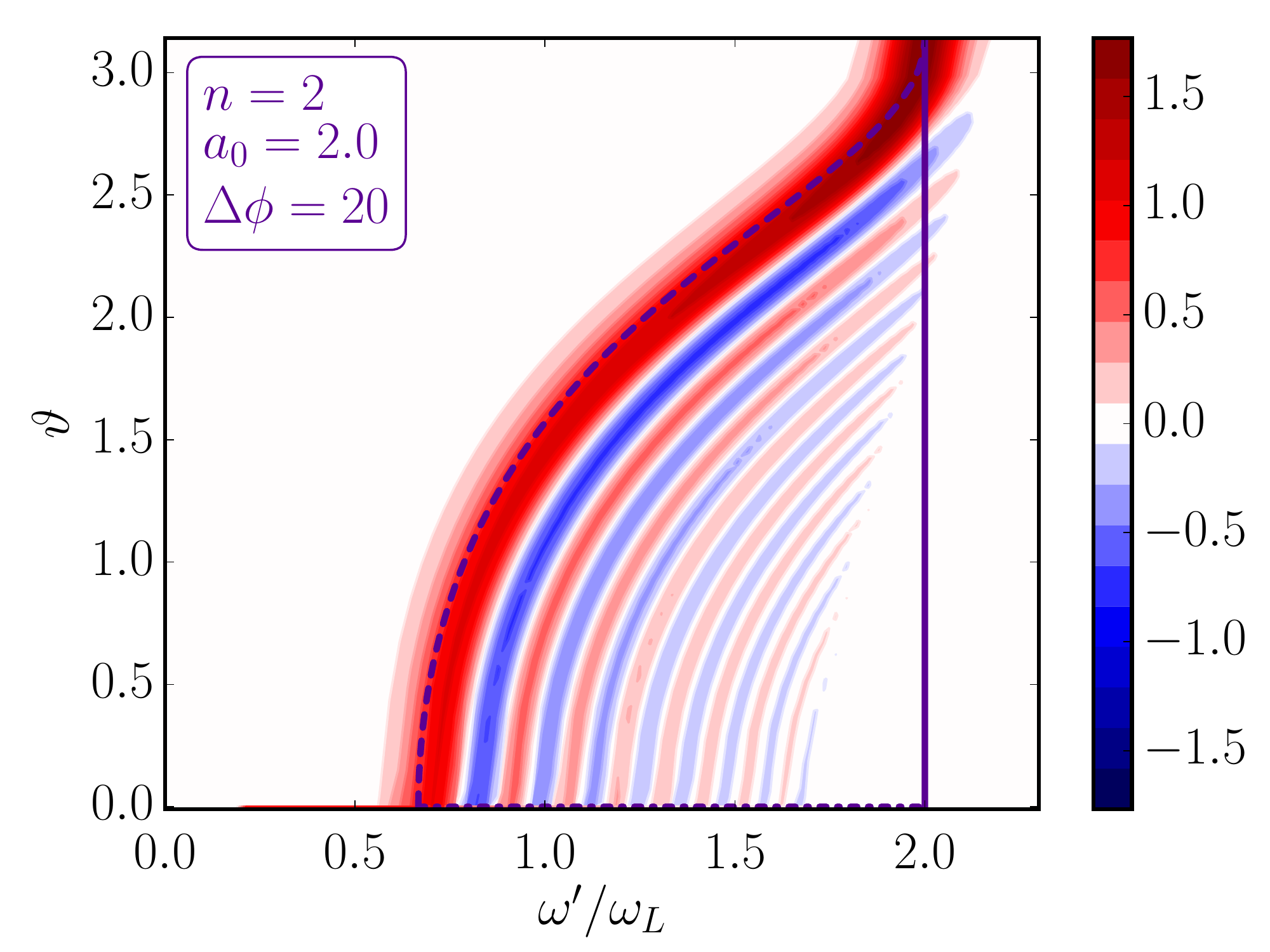}
\includegraphics[width=0.32\columnwidth]{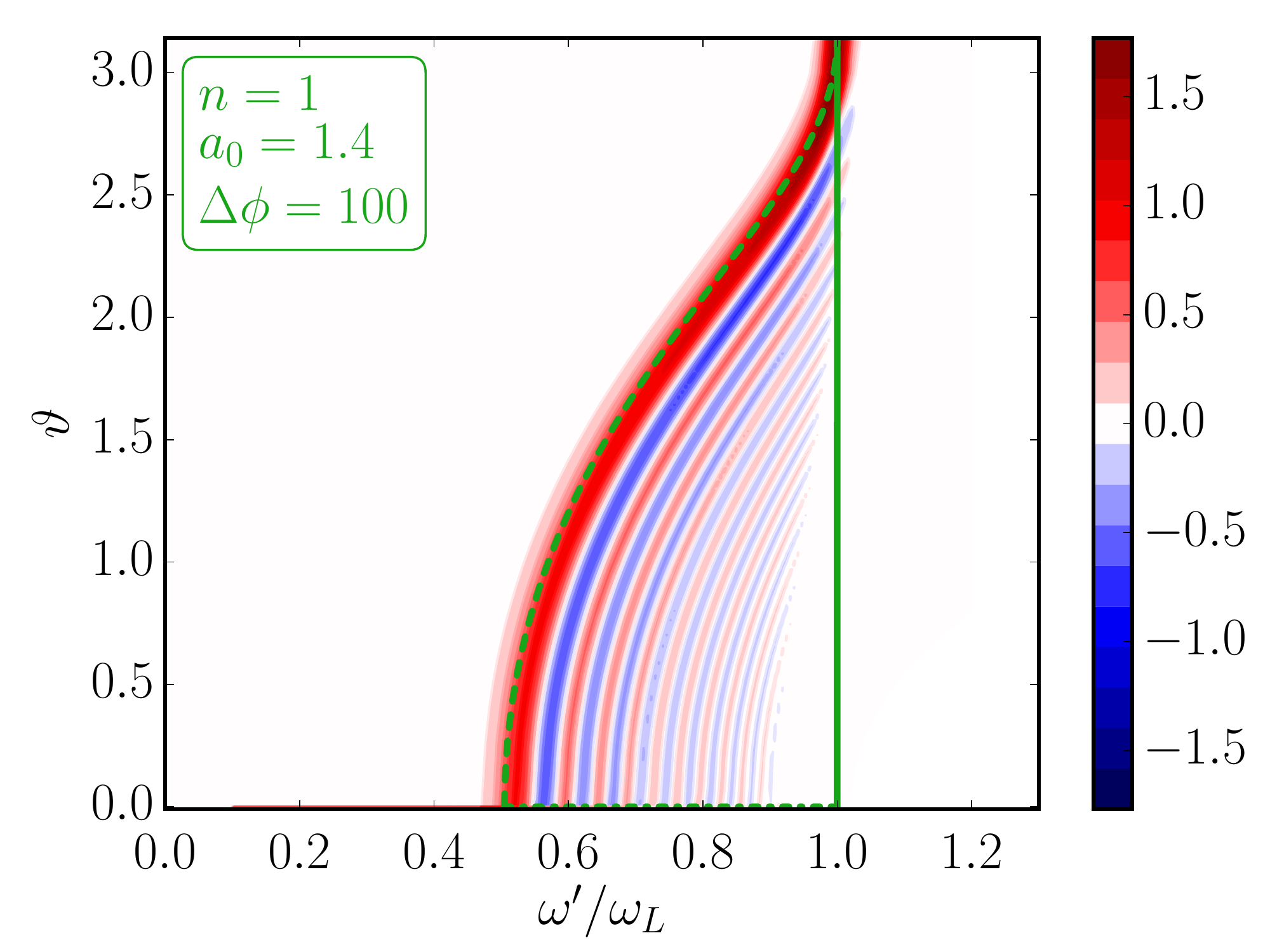}
\end{center}
\caption{{Transformation of the triangular regions in the $\xi$--$\eta$ plane to the
variables $\omega'$ and $\vartheta$. The solid, dashed and dash-dotted curves
correspond to each other.
}
\label{fig:unfolding}}
\end{figure}

{Before we conclude this paragraph, let us explicitly state how the
rather abstract variables $\xi$ and $\eta$ are related to the
physically observable photon frequency $\omega'$ and scattering angle $\vartheta$.
Those expressions explicitly read
\begin{align}
\omega'(\xi,\eta) &= \frac{\xi + n \Delta \phi}{\Delta \phi + \frac{4\omega}{ma_0^2} \, \eta } 
	\label{eq:unfold1} \,,\\
\vartheta (\xi,\eta) &= \arccos \left( \frac{4}{a_0^2} \frac{\eta}{\xi + n\Delta\phi } - 1 \right) \,.
\end{align}

By inspecting Fig.~\ref{fig:cutting} together Eq.~\eqref{eq:def.beta} we find that
for any harmonic $n$ and any value of $r$, the physically relevant
part of the master integral is located in a triangular region with
the corner points $(\xi,\eta)=(0,0)$, $(0,n\Delta \phi \, a_0^2/2)$ and
$(\xi_0,-\xi_0)$ with $\xi_0 = n\Delta \phi \, a_0^2/(2+a_0^2)$.
That means for different $n$, pulse duration $\Delta \phi$ and $a_0$
different parts of the $\xi$--$\eta$ plane describe the spectrum of backscattered photons.
The boundaries of these triangular regions are marked in Fig.~\ref{fig:unfolding} (left panel) for two different sets of parameters. When these areas are transformed to $\omega'$ and $\vartheta$
we obtain the distributions in the centre and right panels, respectively.
They strongly resemble the spectral line shapes in the incident electron rest frame found previously, e.g.~in \citep{Seipt:LasPhys2013}.
}

\begin{figure}
\begin{center}
\includegraphics[width=0.66\columnwidth]{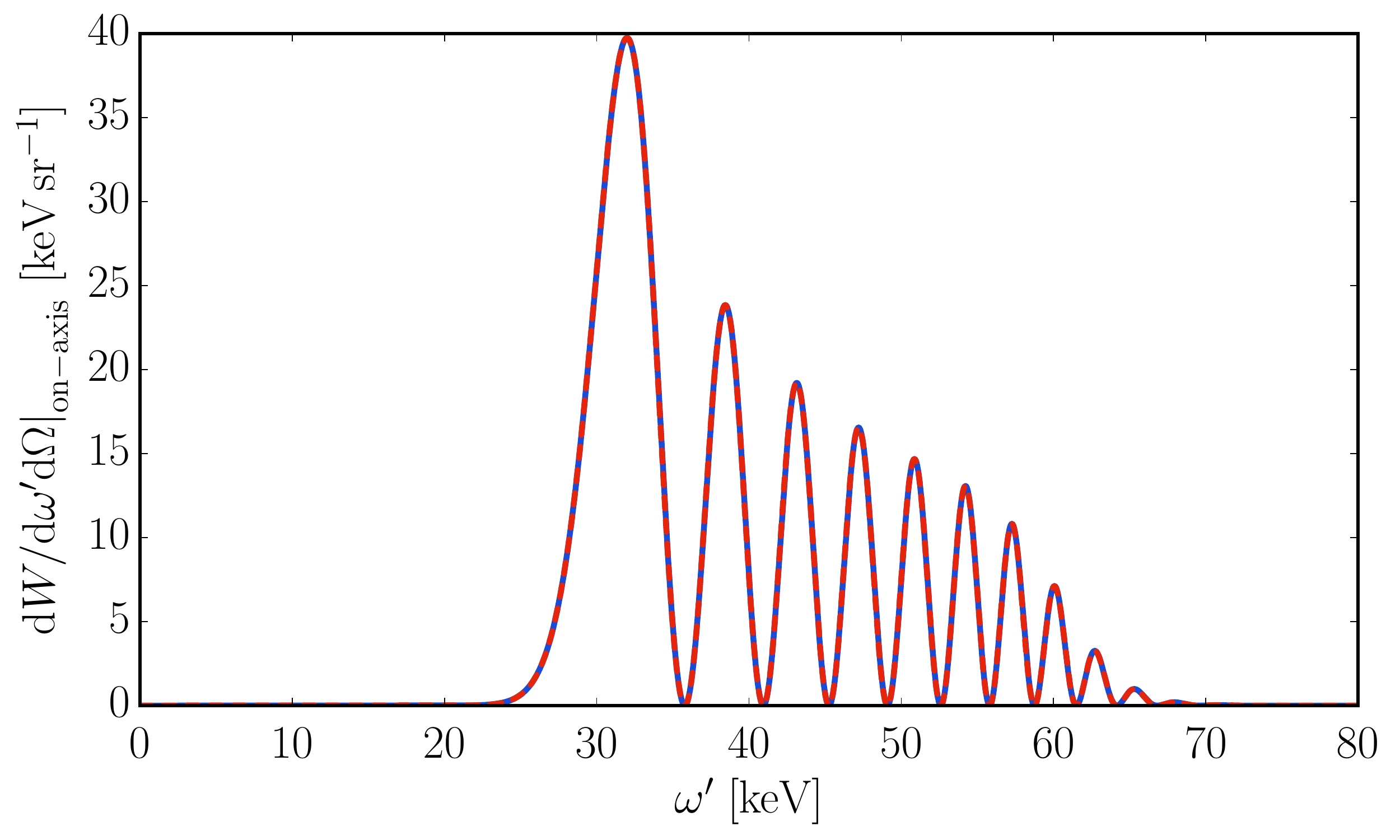}
\end{center}
\caption{{Differential on-axis photon emission probability in the laboratory frame for $a_0=1.5$, $\Delta\phi=25$, a Gaussian pulse shape and $\gamma= 100$. The direct numerical evaluation of the dynamic integrals (blue solid)
coincides perfectly with the evaluations of the master integrals (red dashed).
}
\label{fig:probability}}
\end{figure}

{
For the sake of completeness, we also provide here the corresponding transformation relations
in the laboratory frame where the electron counterpropagates the laser pulse with
a Lorentz factor $\gamma$:
\begin{align}
\omega'_\mathrm{lab} (\xi,\eta) &= 
     \frac{\omega \gamma^2 (1+v)^2 (\xi + n\Delta \phi)}
     { \frac{1+v}{1-v+v \frac{4}{a_0^2}\frac{\eta}{\xi+n\Delta\phi}}
	\left(
	 \Delta \phi + \frac{4}{a_0^2} \frac{\omega \gamma(1+v) \eta }{m}
	\right)     
     }
      \,,\\
\vartheta_\mathrm{lab} (\xi,\eta) &= \arccos \left(
\frac{ \frac{4}{a_0^2} \frac{\eta}{\xi + n\Delta \phi} - 1 +v }{1-v+ v \frac{4}{a_0^2} \frac{\eta}{\xi + n\Delta \phi}}
\right)
 \,, \label{eq:unfold2}
\end{align}
where $v = \sqrt{1 - 1/\gamma^2}$.
The differential on-axis
photon emission probability in the laboratory frame, plotted in Fig.~\ref{fig:probability} for
an initial electron energy of $\unit{51}{\mega\electronvolt}$ ($\gamma=100$) shows perfect agreement between a direct numerical evaluation of the dynamic integrals and the calculation of the master integrals and then transforming from the abstract $\xi$--$\eta$ plane to the photon frequency $\omega'$ and $\vartheta=0$.
}

\subsection{Universal Behaviour of High Order Harmonics}

Let us now draw some conclusions about the shape of very high order harmonics.
One can notice that for large values of $r$, the powers of $g^r(t)$ become strongly localized
around $t\approx 0$.
Thus, the main contribution to the master integral, Eq.~\eqref{eq:def.master.integral}, is provided by the
region of $g(t)$ around the maximum at $t=0$.
For sufficiently smooth pulse envelopes\footnote{This means the first derivative
of the pulse envelope at $t=0$ has to exist.
Due to the symmetry of the pulse envelope it is then equal to zero: $g'(0)=0$.}
$g(t)$ a Taylor expansion around the point $t=0$ reads
\begin{equation}
    g^r(t) \approx \left( 1 - |g''(0)| \frac{t^2}{2} \right)^r  \,.
    \label{eq:high}
\end{equation}
Defining $\tau^2 = 1/|g''(0)|$ and employing the known limit
$(1+x/r)^r\stackrel{r\rightarrow \infty}{\to} \, e^x$, we find
\begin{equation}
    g^r(t) \approx e^{-\frac{r\, t^2}{2\tau^2}}  \,.
\end{equation}
which is of Gaussian shape
and does not depend on any details of the primary pulse shape, except for
the curvature at the maximum, i.e. the second derivative $g''(0)$ at $t=0$.
Note that if one is concerned about laser pulses with a flat top envelope (in the sense that
$g''(0)=0$), the Taylor series in \eqref{eq:high} can be extended further,
and will eventually lead to a Supergaussian shape of $g^r(t)$.

The conclusion is that for sufficiently high harmonic order the master integral
is approximately given by
\begin{align}
\mathcal B_r(\xi,\eta) \approx \intop_{-\infty}^\infty \! \d t \,
e^{-\frac{r}{2}\left(\frac{t}{\tau}\right)^2 + i \xi t + i \eta \int \! \d t \,
g^2(t) } \,.
\end{align}
Because for large values of $r$ the main contributions to this integral come
from a narrow region around $t=0$, we can
approximate $\int dt\, g^2(t)$ polynomially around $t=0$ up to the third order.
This allows us to give an analytic expression for the {master integrals} $\mathcal B_r$
for large values of $r\to \infty$ as
\begin{equation}
    \mathcal{B}_r(\xi,\eta)\approx \frac{2\pi\tau^{2/3}}{\eta^{1/3}}
    \exp\left(-\frac{r}{2\eta}(\xi+\eta)+\frac{r^3}{12(\tau\eta)^2}\right)
    \operatorname{Ai}\left(\frac{r^2}{4(\tau\eta)^{4/3}}
    -\frac{\tau^{2/3}}{\eta^{1/3}}(\xi+\eta)\right)
\end{equation}
in terms of the Airy function Ai \citep{book:Erdelyi}. 
The central conclusion here is that for very large values of $r$ the shape of the harmonic lines
approximate the shape of the harmonics for a Gaussian laser pulse
with temporal duration $\tau / \sqrt{r}= 1/\sqrt{r |g''(0)| } $.
The only specific input from the original pulse shape $g$ that affects the shape of the high-harmonic
spectral lines is the curvature of the pulse envelope at the maximum.

\subsection{Explicit Closed Form Analytic Results for the {Master Integrals}}

\label{sect.explicit}

We finally provide explicit closed form analytic expressions for the {master integrals} $\mathcal B_r(\xi,\eta)$
for several different laser pulse shapes $g$.

\subsubsection{Hyperbolic Secant Pulse Shape}

For a hyperbolic secant pulse, $g(t) = 1/\cosh(t)$, we have $\int g^2 dt = \tanh t$, and the master integral
\begin{align}
\mathcal B_r(\xi,\eta) &= \intop_{-\infty}^\infty \! \d t \, \frac{e^{i\xi t+i \eta \tanh t}}{\cosh^r t} 
\end{align}
can be evaluated by transforming the integration variable according to $z = 1/(2e^t \cosh t)$, yielding
\begin{align}
\mathcal B_r(\xi,\eta) &= 2^{r-1} e^{-i \eta} \intop_{0}^1 \! \d z \,
			\, z^{\frac{r+i\xi}{2}-1} (1-z)^{\frac{r-i\xi}{2}-1}
			\, e^{2i\eta z} \,.
\end{align}
This integral can be evaluated as
\begin{align}
\mathcal B_r(\xi,\eta)		&= \frac{2^{r-1} e^{-i\eta}}{\Gamma(r)}
		 \Gamma\left( \frac{r+i\xi}{2}\right)
		 \Gamma\left( \frac{r-i\xi}{2}\right)
	\:		 _1 F_1\left( \frac{r+i\xi}{2} , r , 2i\eta \right) \,,
	\label{eq:cosh.1F1}
\end{align}
with the Gamma function $\Gamma(z)$ and the confluent hyperbolic function $ _1 F_1 (a,b;z)$
\citep{book:Erdelyi}.
%
An equivalent representation of \eqref{eq:cosh.1F1} can be given in terms of the generalized Laguerre
functions $\mathcal L^\lambda_\nu(z)$ \citep{book:Erdelyi} as
%
\begin{align}
\mathcal B_r(\xi,\eta)		&= \frac{2^{r-1} e^{-i\eta} \pi }{\sin \left( \frac{\pi}{2} (r+i\xi)\right)}
\mathcal L^{r-1}_{-\frac{r+i\xi}{2}} (2i\eta) \,.
	\label{eq:cosh.Laguerre}
\end{align}

\subsubsection{Exponential Pulse Shape}

For a pulse shape of the form $g(t) = e^{-|t|}$ the master integral takes the form
\begin{align}
\mathcal B_r(\xi,\eta) &= \intop_0^\infty \! \d t \, e^{-(r-i\xi)t} \, e^{i \eta e^{-t} \sinh t} \: + \: \rm c.c. \,. 
\end{align}
%
%
After a substitution $z = \frac{i\eta}{2} e^{-2t}$ we get
\begin{align}
\mathcal B_r(\xi,\eta) &= \frac{e^{\frac{i\eta}{2}}}{2} 
\left( \frac{2}{i\eta} \right)^\frac{r-i\xi}{2}
\intop_0^{\frac{i\eta}{2}} \! \d z \, z^{\frac{r-i\xi}{2} - 1} \, e^{-z}  \: + \: \rm c.c. \,,
\end{align}
which can be expressed as
\begin{align}
\mathcal B_r(\xi,\eta) &= \frac{e^{i \eta /2}}{2} \left( \frac{2}{i\eta}\right)^{\frac{r-i\xi}{2}} 
\gamma\left( \frac{r-i\xi}{2}, \frac{i\eta}{2}\right)\: + \: \rm c.c. \,,
\end{align}
with the lower incomplete gamma function $\gamma(a,z)$ \citep{book:Erdelyi},

\subsubsection{Staircase Pulse Shapes}

Let us assume the pulse envelope is staircase, defined as
$g(t) = \sum_{k=1}^N \nu_{k} \chi_{ I_k } ( t ) $ for $t>0$, with $\nu_{k}$ being the height
of the $k$-th step (as measured from the ground level) and the characteristic function
$\chi_{I_k}(t) = 1$ if $t \in I_k= [(k-1)/N,k/N )$, and zero otherwise.
(For $t<0$ the envelope is fully defined by the symmetry $g(-t) = g(t)$.)
For the moment we assume that the step height increases uniformly, $\nu_{k} = (N-k+1)/N$, but a generalization to arbitrary steps is obvious.
A compelling feature of the staircase pulse is
the possibility to approximate many different smooth pulse shapes
in the limit of infinite steps $N\to \infty$, just by adjusting the step heights.
For instance, the uniform staircase discussed here would
converge to a smooth triangle pulse.
%

By splitting the $t$-integration range into the intervals $I_k$ where $g$ is constant
we evaluate the master integral as 
\begin{align}
\mathcal B_r(\xi, \eta) &= 
\sum_{k=1}^N 2 (\nu_{k})^r \:  \frac{\sin \frac{\xi+\eta \, \nu^2_{k}}{2}}{\frac{\xi+\eta \, \nu^2_{k}}{2}} 
\:  \cos \left( \eta \, \Phi_{k} + \frac{2k-1}{N} \frac{\xi+\eta \, \nu^2_{k}}{2} \right)
\end{align}
with 
\begin{align}
\Phi_{k} 
= \frac{1}{N} \sum_{\kappa  = 1}^{k} \nu_{\kappa }^2 -  \frac{k}{N} \nu_{k}^2 \,.
\end{align}
For $N=1$ we recover the well known result of of a sinc profile for the box pulse
which is aligned along the $\eta = -\xi$ diagonal, i.e.~it corresponds to the usual
infinite plane wave red-shift. While the bandwidth of the laser pulse
translates to the Compton scattered light, we see no indication of the
ponderomotive broadening due to a gradual ramp up of the laser intensity. 
How this ponderomotive broadening effect develops can be seen quite instructively when
going to a pulse with more than one step. For $N$ steps we observe
a total of $N$ strips in the $\eta$--$\xi$ plane that are centred along the lines
$\eta = - \xi/\nu_{k}^2$. On each of the steps the radiation is emitted with
with their respective red-shift,
determined by the square of the $k$-th step height as $\ell = n/(1+\beta \nu_{k}^2 )$.
With increasing $N$ these strips eventually are overlapping, reproducing the typical picture
from the smooth pulses discussed before.
Thus, the staircase pulse model discussed here helps to investigate the transition from 
the case of a constant amplitude laser pulse
to the case of smooth pulses where the ponderomotive broadening
sets in and strongly influences the non-linear Compton spectrum.

\section{Conclusions}

In summary, we provided in this paper
a comprehensive and completely analytical evaluation of the
non-linear Compton transition amplitude.
It was found that the dependence on the \textit{shape} of the strong laser pulse
can be traced back to a {class of three-parameter master integrals}.
In addition, all the dependence on the \textit{pulse duration} can be conveniently scaled out from the
master integral.
For certain shapes of the laser pulse envelope we provided explicit analytical expressions for
the {master integrals}. In addition, for very high harmonics we find a universal behaviour of the shape of the harmonic lines.

In this paper we studied only the case of circularly polarized laser light.
The laser polarization affects the form of the Jacobi-Anger type expansion \eqref{eq:Jacobi:Anger}
and the subsequent extraction of the laser pulse envelope from the argument
of the Bessel functions via Eq.~\eqref{eq:BesselJ.multi}.
In the case of an elliptic or linear laser polarization we would encounter generalized two-argument Bessel functions \citep{diss:Seipt,Korsch:JPA2006}.
But eventually the laser pulse shape dependence is described by exactly the same master integrals \eqref{eq:def.Bkr} as for circular laser polarization discussed in this paper.

We would like to stress that the analytical structure of the strong-field $S$ matrix is similar
also for other first-order strong-field QED processes like Breit-Wheeler pair production, or
pair annihilation. Thus, our analytic results could be easily translated to these processes too.

\subsection*{Acknowledgments}

This work was supported in part by the Helmholtz Association (Helmholtz Young Investigators group VH-NG-1037).


\end{document}